\documentclass[twocolumn,showpacs,floatfix,amsmath,amssymb,prb,superscriptaddress]{revtex4}

\usepackage{graphicx}
\usepackage[tight]{subfigure}
\makeatletter

\begin{document} 
\newcommand{\be}{\begin{equation}}
\newcommand{\ee}{  \end{equation}}
\newcommand{\ba}{\begin{eqnarray}}
\newcommand{\ea}{  \end{eqnarray}}
\newcommand{\ve}{\varepsilon}

\title{Conductance of a quantum dot in the Kondo regime connected to dirty wires}

\author{Alberto Camjayi} \affiliation{Departamento de F\a'{i}sica,
  FCEyN and IFIBA, Universidad de Buenos Aires, Pabell\'on 1, Ciudad
  Universitaria, 1428 Buenos Aires, Argentina}

\author{Liliana Arrachea} \affiliation{Departamento de F\a'{i}sica,
  FCEyN and IFIBA, Universidad de Buenos Aires, Pabell\'on 1, Ciudad
  Universitaria, 1428 Buenos Aires, Argentina}

\date{\today}

\begin{abstract}
We study  the transport behavior induced by a small bias voltage through a quantum dot connected to 
one-channel disordered wires by means of a quantum Monte Carlo method. We model the quantum dot 
by the Hubbard-Anderson
impurity and the wires by the one-dimensional Anderson model with diagonal disorder
within a length.  We present a complete description of the probability distribution function of
the conductance within the Kondo regime.
\end{abstract}


\maketitle

\section{Introduction} \label{Intro}

The Kondo effect is one of the most paradigmatic phenomena of electronic correlations. It was proposed long time
ago to explain the peculiar behavior of the resistivity of magnetic impurities in metals as a function of the temperature. \cite{hewson} 
Towards the end of the last century, the interest on this effect was paramount to the mesoscopic community, after it was 
observed in transport experiments in quantum dots.\cite{gold-gor,rev} 

A quantum dot is a confined structure  in contact to metallic wires, where electrons experience a
strong Coulomb repulsion. For temperatures lower than the Kondo temperature $T_K$, the 
Coulomb interaction originates 
an effective coupling between the spin of a localized electron at the dot and the spin of the electrons of the wires.
 The result is the formation of a resonant singlet state  which manifests itself as the opening of a transport channel,
corresponding to a conductance 
$G=e^2/h$
for each spin component. The electrons of the wires that intervene in the formation of these singlets define the so called ``screening cloud,'' which extends over a length $\xi_K= \hbar v_F/(k_B T_K)$, being $v_F$ the Fermi velocity.
The peculiar behavior of $G$ for a dot in the Kondo regime  when the screening cloud does not fit into wires of finite size
has been analyzed in  Refs.~\onlinecite{sim-af, corn-bal}. 
 These studies assume perfect clean wires where the finite-size effects are solely defined by constrictions within the wires at a finite distance from the quantum dot, 
but do not analyze the effect of disorder, which  is an unavoidable ingredient in most of the realistic experimental settings. 

In low dimensional conductors, backscattering induced by disorder produces localization. This  is one of the most dramatic consequences of the quantum coherence characterizing the transport in mesoscopic devices, which enables the interference of the electronic wave function. This effect
has been the subject of many investigations mainly for non-interacting systems,\cite{been, schom, altsh} while there are also some studies for interacting ones.\cite{rusos} In the first case, the probability distribution 
function (PDF) for the conductance at temperature $T=0$,  $P\left(G^0(T=0)\right)$, is analytically known for 1D systems where the disorder is described 
by the disordered Anderson model.\cite{been,dmpk}  That description has been also extended to the case of finite voltage and 
finite temperatures.\cite{mir,moks,vic,us} The function $P\left(G^0(T)\right)$ is completely characterized by a single parameter $ \ell/\xi_{\mathrm{mfp}} = - \langle \ln G^0(0) \rangle$  except for anomalies at the band edges and the band center,\cite{schom, altsh} being $\xi_{\mathrm{mfp}}$ the mean free path, which defines the length beyond which the wave function decays exponentially and $\ell$ the length of the dirty part of the wire. For an interacting impurity, disorder in the environment is expected to affect the development of the Kondo resonance. This problem has been analyzed in the literature under the name of the ``Kondo box.''\cite{delft,kett,muccio,gremcorn}

The interplay between the electronic correlations taking place in the Kondo regime, and the localization induced by the disorder has not been so far considered in the context of the transport properties. In the present work, we precisely  address this important aspect. We identify a crossover in the behavior of $P\left( G(T)\right)$ as the length of the dirty piece of the wire increases over the mean free path $\xi_{\mathrm{mfp}}$ and as the temperature overcomes $T_K$.

\section{Theoretical treatment}

\subsection{Model} \label{model}

 We describe the quantum dot by a Hubbard-Anderson model connected to left ($L$) and right ($R$) wires, which are
dirty within a finite length. The ensuing Hamiltonian is
\be H=H_{d} + \sum_{\alpha=L,R} H_{\alpha}+ H_{cont}, \ee
where the Hamiltonian for the dot includes the effect of the Coulomb repulsion $U$ and the voltage gate $V_g$. It reads
\be
H_{d} =  V_g \sum_{\sigma= \uparrow, \downarrow} n_{d,\sigma} + U n_{d,\uparrow} n_{d, \downarrow}.
\ee
Each wire is modeled as a one-dimensional Anderson Hamiltonian with diagonal disorder within a length 
 $\ell/2=N a$, being $N$ the number of sites of a tight-binding lattice of lattice constant $a$, which we set to be the unit of lenght. The Hamiltonian has the form
  $H_{\alpha}= H_{\alpha,w}+H_{\alpha,c}+H_{\alpha,r}$, being the dirty piece
\be 
H_{ \alpha,w}  =   - t \sum_{j=1}^{N-1}[ c^{\dagger}_{j,{\alpha},\sigma} c_{j+1,{\alpha},\sigma} + H. c. ]+
\sum_{j=1}^{N} \ve_{j, \alpha} c^{\dagger}_{j,{\alpha},\sigma} c_{j,{\alpha},\sigma},
\ee
where the local energies $\ve_{j, \alpha}$ are randomly distributed within a range $[-W,+W]$ for
$1 \leq j \leq N$. The disordered chain is connected through 
\be
H_{\alpha,c}=- t [ c^{\dagger}_{N,\sigma} c_{N+1,\sigma} + H. c. ] 
\ee
to a clean semi-infinite chain (the reservoir) 
\be H_{ \alpha,r}  = 
- t \sum_{j=N+1}^{\infty}[ c^{\dagger}_{j,{\alpha},\sigma} c_{j+1,{\alpha},\sigma} + H. c. ].\ee
The Hamiltonian describing the contact between the dot and the wires reads
\be
H_{cont}=- t \sum_{\alpha=L,R}  [c^{\dagger}_{1,\alpha,\sigma} d_{\sigma} + H. c.].
\ee

\subsection{Conductance} \label{conductance}
As shown in Appendix \ref{apa},
the probability distribution for the ``zero-bias'' conductance through the dot per spin channel, $P\left(G(T)\right)$, in units 
where $e=h=k_B=1$ is evaluated from:
\be \label{cond}
G= \int_{-\infty}^{+\infty} d \omega \overline{\Gamma}(\omega)
\rho_{\sigma} (\omega, T) \frac{ \partial f (\omega)}{\partial \omega},
\ee
where $f(\omega)=[1+e^{\beta (\omega-\mu)}]^{-1}$ is the Fermi function,   with $\overline{\Gamma}(\omega)$ defined in Eq.~(\ref{gamov}) while
 $\rho_{\sigma} (\omega, T)  =  -2 \mbox{Im}[G^R_{0,\sigma}(\omega)]$, is the local density of states (LDS) at the quantum dot.  
 While the non-interacting Green functions 
can be rather straightforwardly evaluated from a recursive procedure, the retarded Green function 
$G^R_{0,\sigma}(\omega)$ depends on the interactions and, thus, on the temperature.

Quantum Monte Carlo methods are powerful  techniques to study the effect of interactions in quantum transport.\cite{qmc, qmcus} For a given disorder realization, we evaluate the Matsubara Green function by means of the quantum Monte Carlo method  of Ref.~\onlinecite{qmcus} and we then use a polynomial fit
to compute this function for real $\omega$.  This procedure is very precise within the low-energy range, $|\omega| <U$. We evaluate
the LDS $\rho_{\sigma}(\omega, T)$ and use this function to compute  $G(T)$. We repeat this procedure for up to $5000$ realizations of disorder.

In the limit of vanishing diagonal disorder $\ve_{l, \alpha}=0$, the model reduces to 
the  Kondo impurity with an effective exchange constant \cite{hewson} 
\be
J= 2 t^2[1/(U+V_g)-1/V_g],
\ee
which
displays a resonance at the Fermi level below the Kondo temperature defined from
\be \frac{2}{J}= \int_{-\infty}^{+\infty}
\frac{\mathrm{d}\omega}{\omega} \tanh\left(\frac{\omega}{2 T_K}\right)\rho_{w}(\omega),\ee
where $\rho_{w}(\omega)= -\mbox{Im}\left[\sum_{\alpha} g_{\alpha}^R(\omega) \right]$ is the density of states of the wires. 
 In a clean system the solution of this equation is $T_K^0$ and the conductance per spin is exactly the conductance quantum for
$T < T_K^0$. The effect of  a mismatching in the contact
 between the wires and the reservoirs,  introduces finite-size 
features in $\rho_w(\omega)$, thus affecting the behavior of the conductance. 
These effects become relevant  within a temperature range of the order of the mean level 
spacing of the wires when this scale is larger than $T_K^0$. This corresponds  to short enough wires for which $\xi_K$
does not fit within the wire, while for longer wires no particular features are observed.\cite{sim-af,corn-bal} 

\section{Results and Discussion} \label{results}

For dirty wires, a given disorder realization
$\ve_{l,\alpha} \neq 0$ defines the behavior of $\rho_w(\omega)$ which in turn determines the corresponding value of $T_K$. 
For an ensemble of disorder realizations, there is a distribution of Kondo temperatures, which in 1D has a crossover from a log-normal distribution centered
at $T_K^0$, to a distribution with  a long tail at low temperatures as the disorder amplitude $W$ increases, while keeping constant the size of the dirty piece.\cite{gremcorn,muccio} 
We have verified that the same behavior is obtained when $W$ is kept fixed and the length of the dirty wires increases. In this case, the crossover takes place as the length of the wires overcomes the mean free path $\xi_{\mathrm{mfp}}$. The mean of the distribution, however, remains approximately $T_K^0$. Consequently, the Kondo screening length is expected to have a probability distribution which 
 does not fit in average within the dirty part of the wires for $\ell < \xi_K$  and the conductance is expected to present a
Kondo behavior distorted by 
finite-size features as a function of temperature. For longer wires, the Kondo cloud fits inside $\ell$, but the disorder becomes more relevant and induces localization. 
In what follows, we analyze the consequence of that crossover in
the behavior of the conductance of this system. For simplicity, we fix $\mu=0$. We also focus on $V_g = U/2$  which
corresponds to the particle-hole symmetric point of the model, at which the Kondo effect is maximum.

\begin{figure}[htb]
\includegraphics[clip,width=\linewidth]{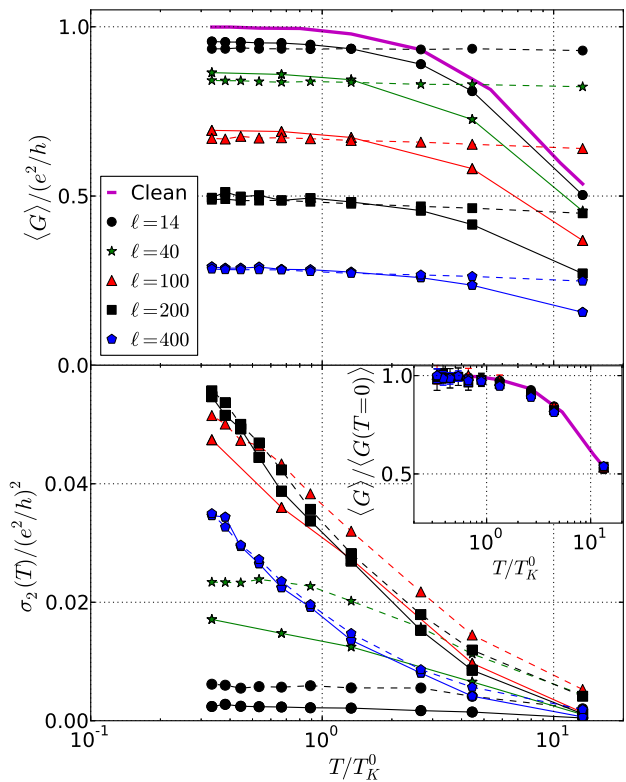}
\caption{(Color online) Top panel: Mean value of the conductance $\langle G(T) \rangle$ as a function of the temperature $T$ for wires of different lengths
$\ell=$ 14, 40, 100, 200 and 400. For comparison, we present in dashed lines
 the corresponding values for a non-interacting dot and 
 in full line the conductance for a dot connected to perfect clean wires.
Bottom panel: Variance $\sigma_2(T)$ for
the same distribution as in the top panel. Inset: Relative conductance $\langle G(T) \rangle/\langle G(T=0) \rangle$. 
The Coulomb interaction is $U=8t$ which, in the clean system corresponds to $T_K^0=0.015t$. 
The disorder strength is $W=t/2$, the gate voltage is $V_g=U/2$ and the chemical potential is $\mu=0$. The mean free path 
of these wires in the non-interacting limit is $\xi_{\mathrm{mfp}}=144$. } \label{fig1}
\end{figure}

The most interesting features introduced by the Kondo effect are expected to emerge at finite temperature. In Fig.~\ref{fig1} we show the mean value of the
PDF of the conductance $\langle G(T) \rangle$ as a function of $T$ for disordered wires of different lengths connected to  an interacting dot,  
along with the corresponding variance $\sigma_2(T) = \langle \left( G(T) -\langle G(T) \rangle \right)^2\rangle$.  The most dramatic feature 
noticed in this figure is the exponential drop of the mean conductance $\langle G(T) \rangle$ as the temperature grows above the Kondo
temperature of the clean system $T_K^0$. Although the absolute drop is more pronounced for short chains with $\ell < \xi_{\mathrm{mfp}}$, the relative mean values
$\langle G(T) \rangle/\langle G(T=0) \rangle$ coincide with the conductance for the clean system shown in full lines for comparison (see inset in Fig.~\ref{fig1}).
This is in strike contrast with the behavior of the mean conductance for $U=0$,                  
$ \langle G^0(T) \rangle$,
also shown in the Fig.~\ref{fig1} (see dashed lines), which remains approximately
constant within a range of temperatures of the order of the bandwidth of the reservoirs.\cite{us}
 In the clean system,  it is known that the process of singlet formation with one spin at the dot and one spin at the wires producing the Kondo resonance is effective only at low temperatures $T<T_K^0$. Above the
Kondo temperature the dominating process is the Coulomb blockade, which is a consequence of the high energy cost to introduce
an additional electron in the dot, once it has been  previously occupied by another one. 
In this regime, the Kondo peak at the Fermi energy of the LDS at the dot melts into a valley which lies between two peaks separated by $U$,\cite{hewson} and the consequence is a dramatic drop in the conductance. For dirty wires  in which case we have a  distribution of Kondo temperatures, we can also identify such processes in  the drop of the conductance as a temperature-driven transition from the Kondo to the Coulomb blockade regimes. For short wires with
$\ell<\xi_K$ the finite-size features typically observed at $T \sim 8t/\ell$,\cite{sim-af,corn-bal} depend on the particular realization of disorder and are
not appreciated in the average.

A more subtle feature to analyze is the difference between the interacting and non-interacting PDF, which takes
place for short wires and $T<T_K^0$. Although small, this difference can be appreciated in the Fig.~\ref{fig1} in the mean value and in   the
variance $\sigma_2(T)$ relative to the corresponding non-interacting value $\sigma_2^0(T)$. Remarkably, for short wires,  $\langle G(T) \rangle > \langle G^0(T) \rangle $, while  $\sigma_2(T)$ is a decreasing function of $T$, as the non-interacting one,  although systematically smaller.

\begin{figure}[htb]
\includegraphics[clip,width=1\linewidth]{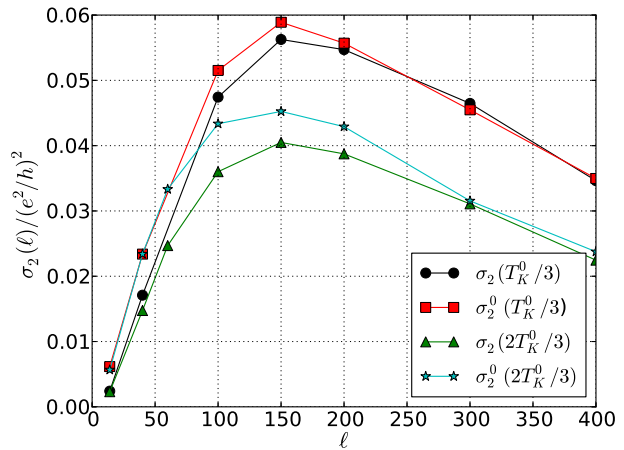}
\caption{(Color online) Variance for the interacting ($\sigma_2(T)$) and non-interacting ($\sigma_2^0(T)$) system as 
a function of the length $\ell$ for wires at low temperature $T \approx T_K^0/3 \mbox{ and } 2T_K^0/3$. 
Other parameters are the same as in Fig.~\ref{fig1}. } 
\label{fig2}
\end{figure}

\begin{figure*}[htb]
\subfigure{\includegraphics[clip,width=0.3\linewidth]{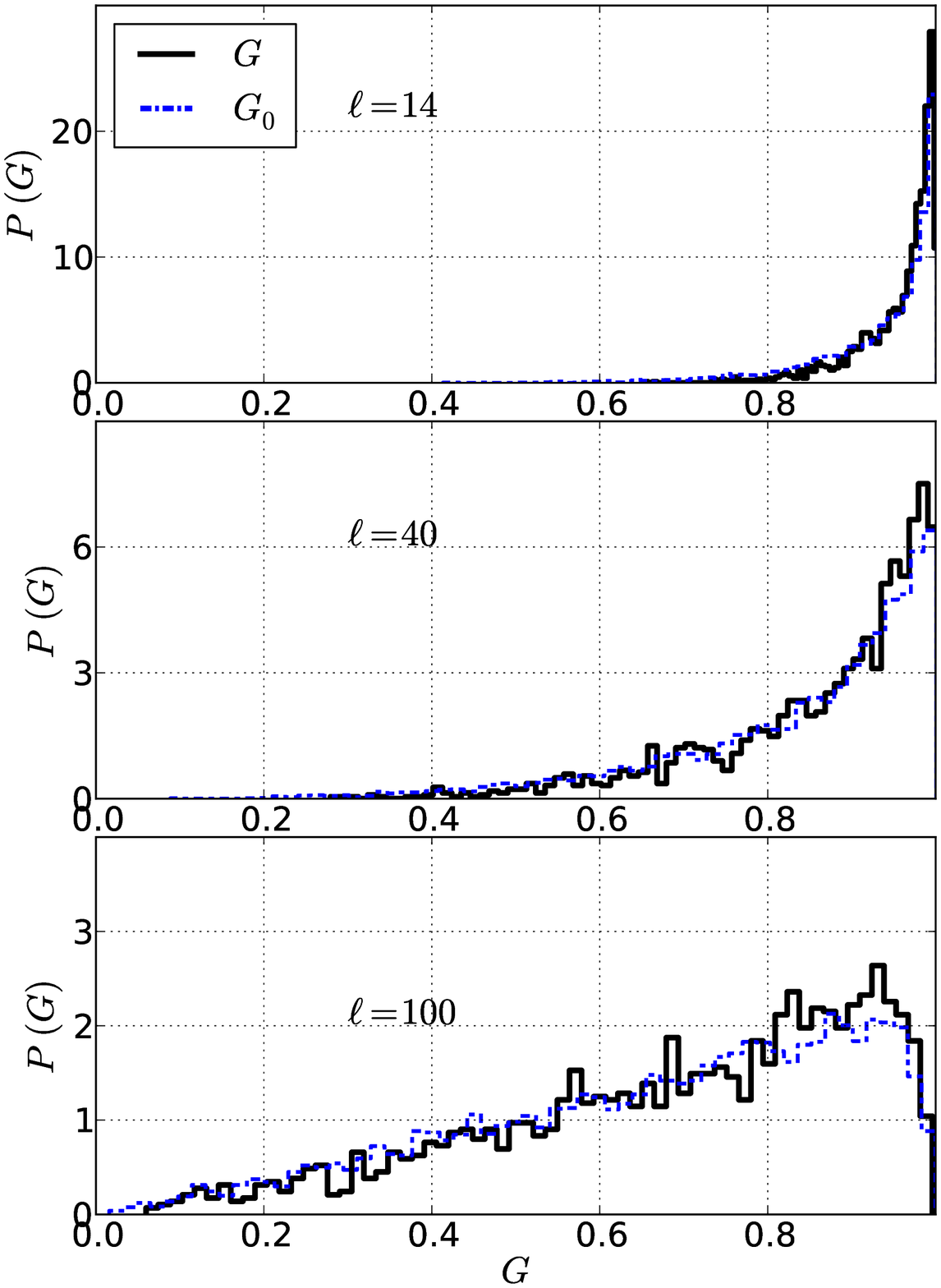}}
\subfigure{\includegraphics[clip,width=0.3\linewidth]{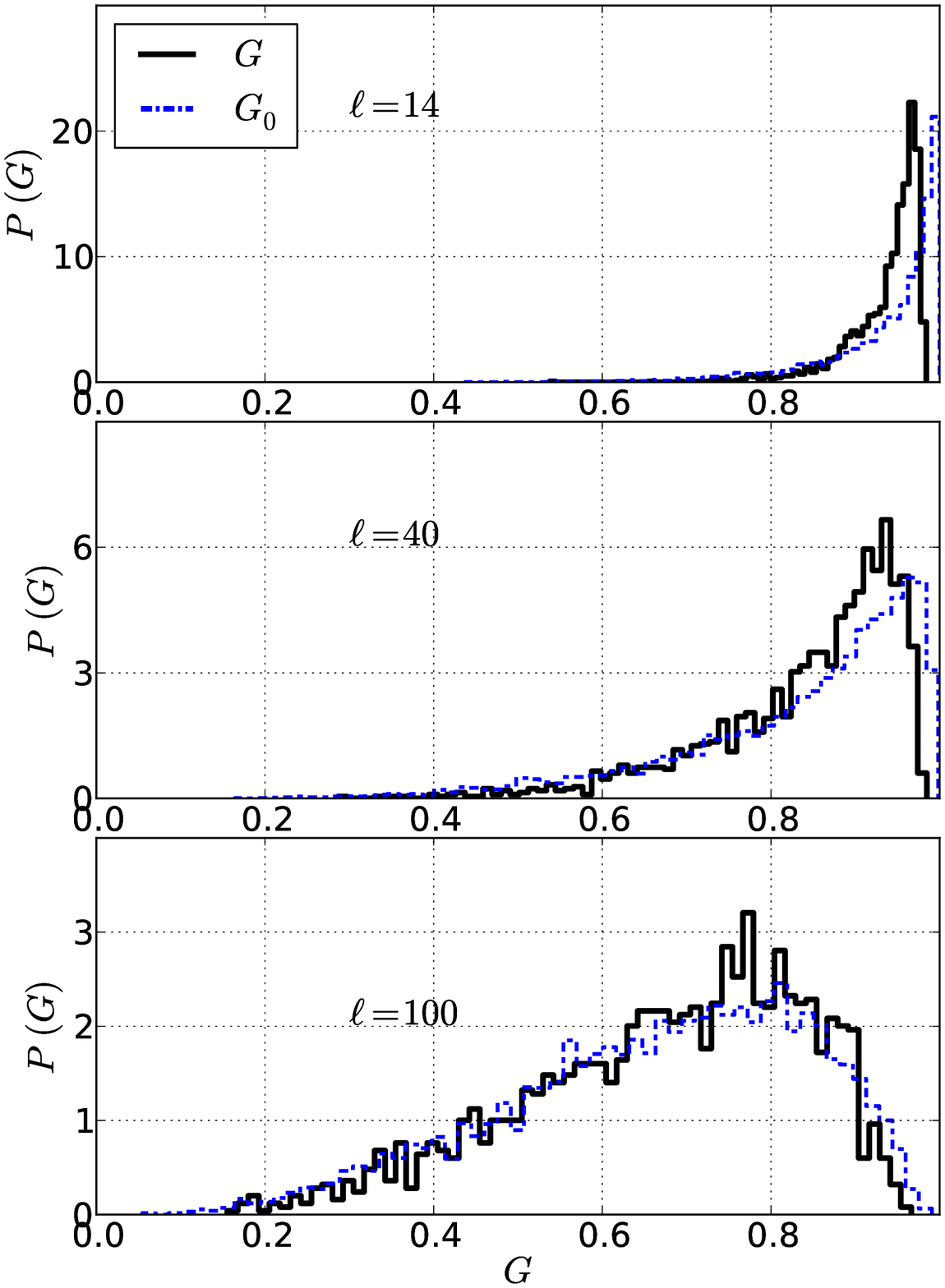}}
\subfigure{\includegraphics[clip,width=0.3\linewidth]{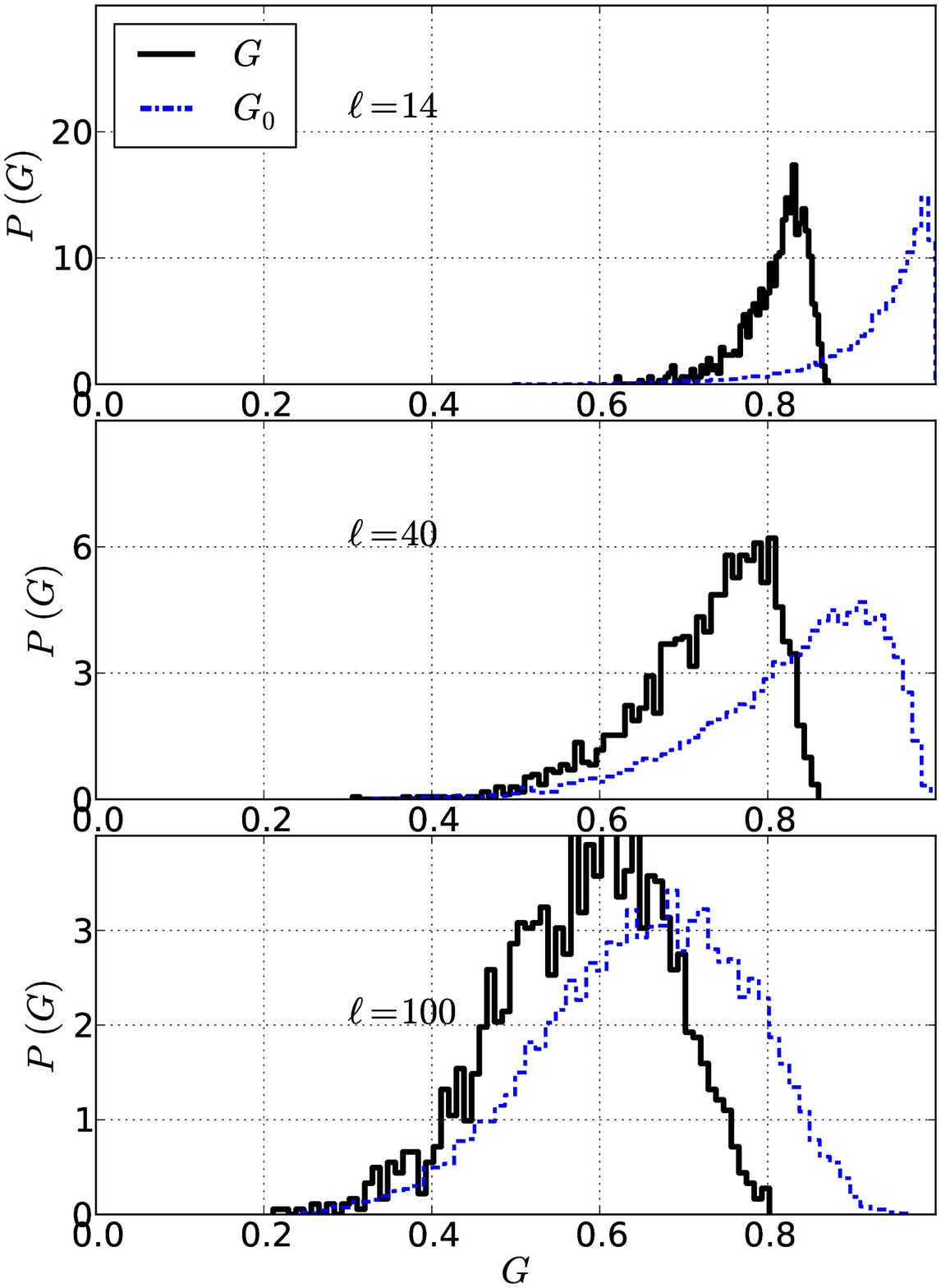}}
\caption{(Color online) PDF for the full $P\left(G \right)$ (full line) and non-interacting $P\left( G_0 \right)$ (dashed-dotted line) conductance. 
Left, central and right panels correspond to $T = T_K^0/3$ , $T = 1.33T_K^0$ and $T \approx 4.5T_K^0$ respectively. On each panel
three different lengths of the dirty wires $\ell=14, 40 \mbox{ and } 100$ are shown. Other parameters are the same as in Figs.~\ref{fig1} and \ref{fig2}. } \label{fig3}
\end{figure*}

In Fig.~\ref{fig2} we focus on the behavior of the variance of both an interacting and a non-interacting dot at a fixed temperature $T<T_K^0$  as a function of the length $\ell$.  
It is important to mention that the temperature is low enough to allow us to express the conductance by $G(T)= \overline{\Gamma}(0) \rho_{\sigma}(0, T)$, i.e. to disregard the corrections of approximating the function $\partial f/\partial\omega \sim \delta (\omega)$. 
It is clear that the low temperature difference between the interacting and non-interacting 
variance of the PDF persists up to the mean free path $\xi_{\mathrm{mfp}}$. Actually, it can be seen that 
for $\ell>\xi_{\mathrm{mfp}}$, the full conductance distribution function exactly coincides with the non-interacting log-normal distribution, 
$P\left( G(T)\right)=P\left( G^0(T) \right)$. This it is shown in Fig.~\ref{fig3} for several temperatures, from low $T = T_K^0/3$, where the differences are small up to $T \approx 4.5T_K^0$, where the difference between both PDF is evident. 

In order to gain insight on the low $T<T_K^0$ behavior of $P\left( G(T) \right)$ as a function of $\ell$, we analyze the behavior of the PDF
for the LDS at the Fermi energy of the dot, $P\left( \rho_{\sigma}(0, T) \right)$. To this end we
resort to the Fermi liquid description, expected to be valid within this regime.
This implies that for low $\omega$ and $T$ the self-energy representing the many-body interactions at the dot, $\Sigma_{\sigma}(\omega, T) = \Sigma_{\sigma}^{\prime}(\omega, T)+
i \Sigma_{\sigma}^{\prime \prime} (\omega, T)$,  behaves
as  $\Sigma_{\sigma}^{\prime \prime} (\omega, T) \propto \omega^2 + \pi^2 T^2$.\cite{oguri} In that limit, 
the Green function for the impurity can be approximated by
\be \label{fermi}
G^R_{d,\sigma}(\omega) = \frac{z (T) }{\omega - z(T)  \overline{V_g}-i z (T) \left( \Sigma^{\prime \prime}_0(\omega)
+ \Sigma^{\prime \prime}_{\sigma}(\omega,T) \right)},
\ee
where $z$ is the quasiparticle weight, $z^{-1}(T) = 1-\partial_{\omega} \Sigma_{\sigma}^{\prime}(0,T)$, and
$\overline{V_g}=  V_g-\Sigma_0^{\prime}(0)-\Sigma_{\sigma}^{\prime}(0,T)$, being $\Sigma_0(\omega)=
 t^2 \sum_{\alpha} g_{\alpha}^R(\omega)$. Since we focus on $V_g = U/2$ then $\overline{V_g} \approx \Sigma_0^{\prime}(0)$. From this expression it can
be verified that,
up to ${\cal O}(T^2)$,
the Fermi liquid  LDS at the quantum dot can be written as 
\be\label{rhofer}
\rho^{\mathrm{Fl}}_{\sigma}(0, T)= \rho^0(0)\left[1+\frac{\Sigma^{\prime 2}_0(0)-\Sigma^{\prime \prime 2}_0(0)}{\Sigma^{\prime 2}_0(0)+\Sigma^{\prime \prime 2}_0(0)}
\frac{\Sigma^{\prime \prime}_{\sigma}(0,T)}{\Sigma^{\prime \prime}_0(0)}\right],
\ee
being $\rho_0(\omega)$ the non-interacting LDS, while $\Sigma^{\prime \prime}_{\sigma}(0,T)$ represents the inelastic
scattering effects introduced by the Coulomb interaction that take place at finite $T$.

\begin{figure*}[htb]
\subfigure{\includegraphics[clip,width=0.4\linewidth]{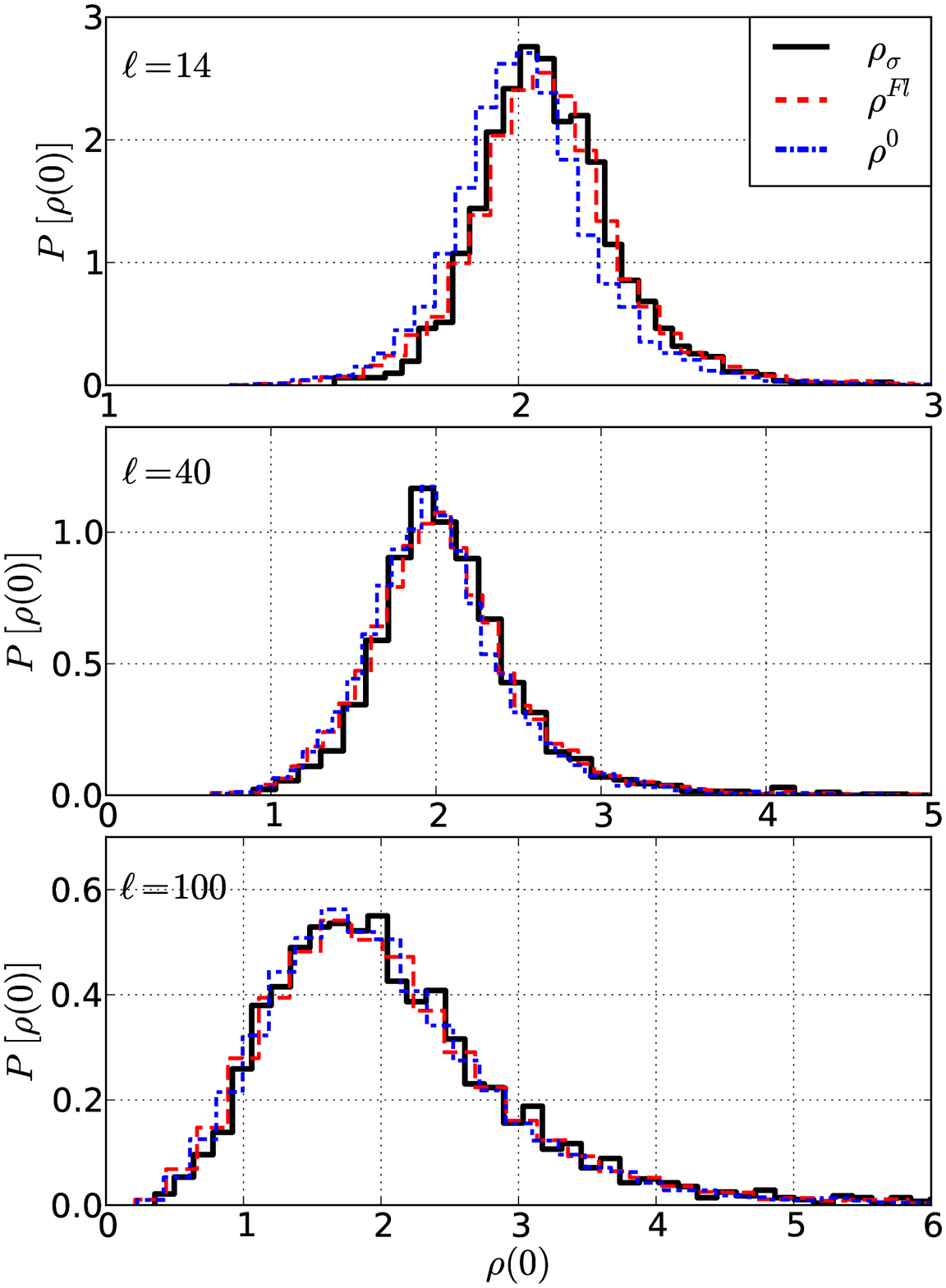}}
\subfigure{\includegraphics[clip,width=0.4\linewidth]{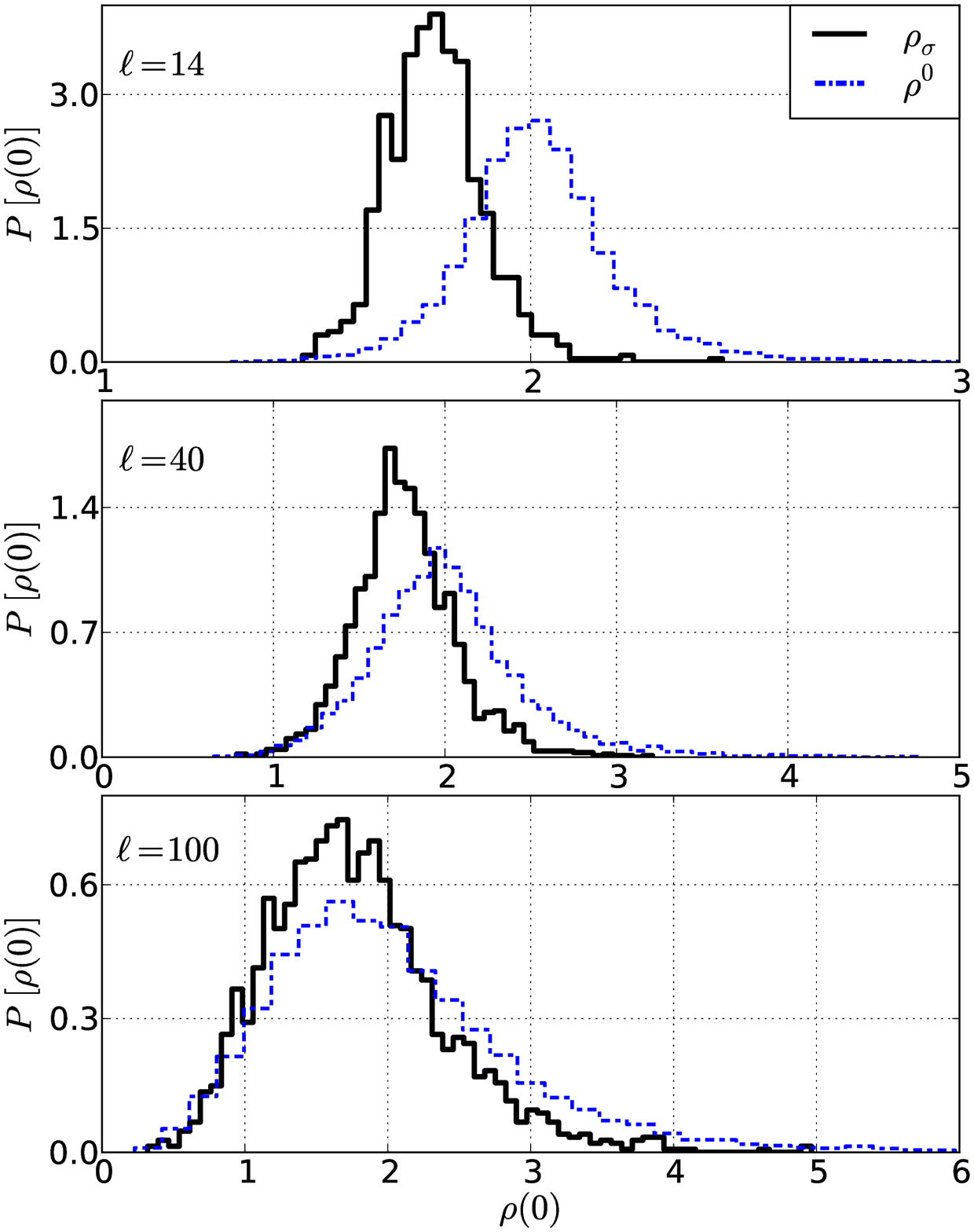}}
\caption{(Color online) PDF for the LDS, $P\left(\rho_{\sigma}(0)\right)$ at $T= T_K^0/3$ (left panel) and $T \approx 4.5T_K^0$ 
(right panel) of an interacting quantum dot in the Kondo regime, corresponding
to three different lengths of the dirty wires $\ell=14, 40 \mbox{ and } 100$. For comparison, we also show in dashed-dot lines the function $P\left( \rho^0(0) \right)$ corresponding to the non-interacting system. For the lower temperature, we also show in dashed lines the PDF corresponding to the Fermi-liquid description. Other parameters are the same as in Figs.~\ref{fig1} and \ref{fig2}. } \label{fig4}
\end{figure*}

The Monte Carlo procedure we follow allows us to evaluate, for a given disorder realization, 
$\rho_{\sigma}(\omega, T)$, while $\rho^0(0)$ can be obtained independently by a numerically exact recursive procedure. 
In Fig.~\ref{fig4} we show the histograms corresponding to these two LDS and it is clear that these two distributions are different for
short wires with $\ell < \xi_{\mathrm{mfp}}$, which justifies the difference in the first two moments of $P \left( G(T) \right)$. 
The quantum Monte Carlo data also provides the distribution for $\Sigma_{\sigma}(0,T)$. The imaginary part of this 
self-energy is peaked around its mean value, which we consider along with  the data for $\rho^0(0)$ and $\Sigma_0$ to evaluate the Fermi liquid
density of states given by Eq.~(\ref{rhofer}). The corresponding histograms, presented in Fig.~\ref{fig4}, show that
$P\left(\rho^{\mathrm{Fl}}_{\sigma}(0, T)\right)$ reproduces very well  the exact $P\left( \rho_{\sigma}(0, T) \right)$ within the whole range
of lengths. The fact that for sufficiently long chains the distribution function $P\left( \rho_{\sigma}(0, T) \right)$ along with the Fermi liquid approximation 
exactly coincides with the corresponding distribution for the non-interacting case can be understood within the Fermi-liquid description. Indeed, an analysis of the mean value of $\Sigma^{\prime \prime}_{\sigma}(0,T)$ as a function of the length indicates that this quantity 
decreases linearly as $\ell$ increases and falls to zero at $\ell \sim \xi_{\mathrm{mfp}}$. This is consistent with a picture where
the inelastic scattering processes accounted by $\Sigma^{\prime \prime}$ are determined by the 
square of the available phase space, which  is $\propto T^2$ in a clean system and 
reduces as the system becomes dominated by the disorder and localizes. 

 Substituting the Fermi liquid density of states  (\ref{rhofer}) in (\ref{cond}) defines a probability distribution which reproduces the 
 corresponding function for the conductance $P(G(T))$ for $T < T_K$. The particular case of
$T=0$ is rather straightforward after noticing that $\rho^{\mathrm{Fl}}_{\sigma}(0,0) \equiv \rho^0(0)$, which means that
 the conductance distribution for the interacting dot  with dirty wires in the Kondo regime exactly coincides in this limit with the corresponding to
  the non-interacting case, which is analytically known.\cite{been,dmpk,schom,altsh}  

For completeness and for comparisson, we present in the right panels of Fig.~\ref{fig4} the
histograms corresponding to the LDS at the chemical potential for a high temperature, larger than
$T_K^0$. In this case, the interacting PDF difers significantly from the corresponding one for the
non-interacting system, mainly for short wires with lengths below $\xi_{\mathrm{mfp}}$. In addition we failed to reproduce these histograms with a Fermi liquid model like the
one defined by  Eq.~(\ref{rhofer}). That is rather expected, since this temperature falls within the 
Coulomb blockade regime of the clean system, where the Fermi liquid description breaks down.

\section{Conclusion}

To conclude, the finite size of the wires has been identified as a source of anomalous behavior  in the transport properties of quantum dots, particularly within the Kondo regime.\cite{sim-af, corn-bal} In the present contribution we have added the ingredient of disorder to that scenario.
We were able to propose a full description for the PDF of the conductance of a quantum dot connected to disordered wires within the Kondo regime. For $T=0$ this function coincides with the one for non-interacting systems. 
For $0<T<T_K^0$ and $\ell< \xi_{\mathrm{mfp}}$, it can be described  by a Fermi liquid density of states corresponding to Eq.~(\ref{rhofer}), while for $\ell> \xi_{\mathrm{mfp}}$ there is a crossover to a regime where the  localization dominates over the inelastic scattering effects originated in the interactions and 
the PDF corresponds to the same log-normal  function of the non-interacting system. For increasing temperatures $T>T^0_K$, the  exponential drop of the conductance that characterizes the Kondo regime is also observed in these systems. The drop is milder as the length of the dirty wires
increases, since the corresponding values for the low-temperature regime are significantly smaller than the conductance quantum.

\section{Acknowledgments}We thank A. Aligia and C. Balseiro for useful discussions. 
This work is supported by CONICET, ANCyT and UBACYT, Argentina. We also thank Ibercivis.es for computing facilities.

\appendix

\section{Derivation of the expression for the conductance} \label{apa}
We briefly present the main steps leading to the expression
for the conductance of Eq.~(\ref{cond}). 
 These are basically the same  
presented in Ref.~\onlinecite{meiwi}. We also focus on the so called ``zero bias'' regime, where the potential difference $V$ is assumed to be 
infinitesimally small.

For a given disorder realization the current per spin channel through the contact between the lead $\alpha$ and the quantum dot, reads
\be \label{cur}
J_{\alpha}=\frac{- 2 e t}{\hbar} \int \frac{d \omega}{2 \pi}\mbox{Re}[G^<_{\alpha,0, \sigma}(\omega)],
\ee
being $G^<_{\alpha,0, \sigma}(\omega)$ the Fourier transform of the lesser Green's function 
$G^{<}_{\alpha,0, \sigma}(t,t^{\prime})= i \langle c^{\dagger}_{1 \alpha, \sigma} (t) d_{\sigma} (t^{\prime}) \rangle$. 
The latter
obeys the following Dyson's equation
\be
G^<_{ \alpha,0, \sigma}(\omega) = - t [g^<_{\alpha,\sigma}(\omega) G^A_{0,\sigma}(\omega)+ g^R_{\alpha,\sigma}(\omega) G^<_{0,\sigma}(\omega)],
\ee
where $g^R_{\alpha}(\omega)$ is the retarded Green's function corresponding to the wire uncoupled from the dot, with the two spatial coordinates at 
 the first site $1 \alpha$ of the lead $\alpha$,
while $g^<_{\alpha,\sigma}(\omega) = -i f_{\alpha}(\omega) 2 \mbox{Im}[g^R_{\alpha}(\omega)]$, with the Fermi function $f_{\alpha}(\omega)= 
[e^{(\omega -\mu_{\alpha})/T_{\alpha}}+1]^{-1}$, is the corresponding  lesser Green's function. $G^A_{0,\sigma}(\omega)=[G^R_{0,\sigma}(\omega)]^*$ is the
advanced Green's function of the dot connected to the wires, with the two spacial coordinates at the dot, while $G^<_{0,\sigma}(\omega)$ is the corresponding
lesser Green's function.

Substituting in (\ref{cur}), the probability distribution function  for this current over the ensemble of disorder realization cast 
\begin{multline}
P\left[ J_{\alpha}\right] = \frac{e}{\hbar} \int \frac{d \omega}{2 \pi} \left\{ P \left[\Gamma_{\alpha}(\omega) \rho_{\sigma}(\omega ) \right] f_{\alpha}(\omega) \right . \\
- P \left . \left[\mbox{Re}\left( \Sigma^R_{\alpha}(\omega) G^<_{0,\sigma}(\omega) \right) \right] \right\} ,
\end{multline}
being
\begin{equation}
\Sigma^R_{\alpha}(\omega)= t^2 g^R_{\alpha}(\omega), \label{sig}
\end{equation}
where $\Gamma_{\alpha}(\omega)= - 2 \mbox{Im}[\Sigma^R_{\alpha}(\omega)]$ is the spectral function representing the hybridization to the reservoir $\alpha$ and
$\rho_{\sigma}(\omega )= -2 \mbox{Im}[G^R_{0,\sigma}(\omega)]$ for a given disorder
realization. The  spectral function of the reservoir
can be in general expressed as $\Gamma_{\alpha}(\omega) = 2 \pi t^2 \sum_{k_{\alpha}} |v_{1,k_{\alpha}}|^2 \delta(\omega- \varepsilon_{k_{\alpha}})$, where
$c_{l,\alpha,\sigma}= \sum_{k_{\alpha}}v_{l,k_{\alpha}} c_{k_{\alpha},\sigma}$ defines the change of basis that diagonalizes the Hamiltonian
$H_{\alpha}= H_{\alpha,w}+H_{\alpha,}+ H_{\alpha,r}$. However, the most convenient way to evaluate this function is by first evaluating  $g^R_{\alpha}(\omega)$  by 
recourse to a decimation procedure and then using (\ref{sig}). Instead, $\rho_{\sigma}(\omega )$ depends on the interacting Green
function of the dot, which must be calculated with QMC.

For uncorrelated local disorder, the probability distribution function 
is symmetric under left-right inversion, thus
 $P \left[ \mbox{Re}[\Sigma^R_{L}(\omega) G^<_{0,\sigma}(\omega)] \right] = P \left[ \mbox{Re}[\Sigma^R_{R}(\omega) G^<_{0,\sigma}(\omega)] \right]$
 and $P \left[ \Gamma_{L}(\omega) \rho_{\sigma}(\omega ) \right] = P \left[\Gamma_{R}(\omega) \rho_{\sigma}(\omega ) \right]$. Taking into account
 the conservation of the current $J_L=-J_R =J$, using the assumption of small bias $V$ and considering that the two leads are at the same temperature $T$ we can
 write the following expression for the probability distribution function for the conductance
 \be
 P\left[ G \right]=   \frac{ e}{\hbar} \int \frac{d \omega}{2 \pi} P \left[ \overline{\Gamma}(\omega) \rho_{\sigma}(T,\omega ) \right]
\frac{ \partial f(\omega)}{\partial \omega},
\ee 
being 
\be \label{gamov}
\overline{\Gamma}(\omega)= \frac{\Gamma_{L}(\omega) \Gamma_R(\omega)}{\Gamma_{L}(\omega) + \Gamma_R(\omega)},
\ee
while $\rho_{\sigma}(T,\omega )$ is the equilibrium density of states of the dot, in contact to the reservoirs at the same chemical potential $\mu$ and
temperature $T$. Notice that this procedure is valid only for small bias voltage $V$ and corresponds
 to an evaluation of the current which is exact only up to $O(V)$. For higher voltages the drop along the dot and or the wires (depending on where
 the bias voltage is assumed to be applied) should be taken
into account. In addition, a full non-equilibrium evaluation of $\rho_{\sigma}(\omega)$ is necessary. We have verified that in the limit of 
a non-interacting dot, where the density of states $\rho_{\sigma}(\omega)$ can be exactly evaluated, we recover the probability distribution function for 
the conductance of Refs.~\onlinecite{been, us}. The latter corresponds to the histograms in dashed lines of Fig.~\ref{fig3}.




\begin{thebibliography}{11}
\bibitem{hewson} A.C. Hewson, {\em ``The Kondo problem to heavy fermions,''} (Cambridge Univ. Press, Cambridge, 1993).
\bibitem{gold-gor} D. Goldhaber-Gordon, H, Shtrikman, D. Mahalu, D. Abusch-Magder, U. Meirav, and   M. A. Kastner,
 Nature {\bf 391}, 156 (1998).
\bibitem{rev}For a review see L.I. Glazman, M. Pustilnik,  in ``Nanophysics: Coherence and Transport,'' eds. H. Bouchiat et al. 
(Elsevier, 2005), pp. 427-478 (2005) and references therein.
\bibitem{sim-af}I. Affleck and P. Simon, Phys. Rev. Lett {\bf 86}, 2854 (2001); P. Simon and I. Affleck {\bf 89}, 206602 (2002).
\bibitem{corn-bal} P. S. Cornaglia and C. A. Balseiro, Phys. Rev. Lett. {\bf 90},  216801 (2003).
\bibitem{been}C. W. J. Beenakker, Rev. Mod. Phys. {\bf  69}, 731 (1997) and references therein.
\bibitem{rusos}D.M. Basko,  I. Aleiner and B. Altshuler, Annals of Physics {\bf 321}, 1126 (2006); 
I. V. Gornyi, A. D. Mirlin and D. G. Polyakov, Phys. Rev. B {\bf 75}, 085421 (2007)
\bibitem{dmpk}P. A. Mello and N. Kumar, {\em ``Quantum Transport in Mesoscopic Systems. Complexity and statistical fluctuations,''}
Oxford University Press, Oxford, (2004).
\bibitem{mir}I. V. Gornyi, A. D. Mirlin and D. G. Polyakov, Phys. Rev. Lett. {\bf 95}, 206603 (2005).
\bibitem{moks}M. Mo\^sko, P. Vagner, M. Bajdich and T. Sch\"apers, Phys. Rev. Lett. {\bf 91}, 136803 (2003).
\bibitem{vic}V. A. Gopar and P. W\"oelfle, Europhys. Lett. {\bf 71}, 966 (2005).
\bibitem{us}F. Foieri, M. J. S\'anchez, L. Arrachea and V. A. Gopar, Phys. Rev. B {\bf 74 }, 165313 (2006). 
\bibitem{schom}H. Schomerus and M. Titov, Phys. Rev. B {\bf 67}, 100201 (2003). 
\bibitem{altsh}L. I. Deych, M. V. Erementchouk, A. A. Lisyansky and B. L. Altshuler, Phys. Rev. Lett. {\bf 91}, 096601 (2003).
\bibitem{delft}W. B. Thimm, J. Kroha and J. von Delft, Phys. Rev. Lett. {\bf 82}, 2143 (1999).
\bibitem{kett}A. Zhuravlev, I. Zharekeshev, E. Gorelov, A. I. Lichtenstein, E. R. Mucciolo and S. Kettemann, Phys. Rev. Lett. {\bf 99}, 247202 (2007).
\bibitem{muccio}S. Kettemann and E. R. Mucciolo, Phys. Rev. B {\bf 75}, 184407 (2007); 
T. Micklitz, T. A. Costi and A. Rosch,  ibid. B {\bf 75}, 054406 (2007).
\bibitem{gremcorn}P. S. Cornaglia, D. R. Grempel and C. A. Balseiro, Phys. Rev. Lett., {\bf 96}, 117209 (2006).
\bibitem{meiwi}Y. Meir and N. S. Wingreen, Phys. Rev. Lett. {\bf 68}, 2512 (1992).
\bibitem{qmc}L. Arrachea and M. J. Rozenberg, Phys. Rev. B {\bf 72}, 041301 (2005); E. Gull, A. J. Millis, A. I. Lichtenstein,
A. N. Rubtsov, M. Troyer and P. Werner,  Rev. Mod. Phys. {\bf 83}, 349 (2011); 
R. H\"utzen, S. Weiss, M. Thorwart and R. Egger, Phys. Rev. B {\bf 85}, 121408 (2012); K. F. Albrecht, {\em et al}, arXiv:1206.4464.
\bibitem{qmcus}P. Werner, A. Comanac, L. de Medici, M. Troyer and A. J. Millis, Phys. Rev. Lett. {\bf 97}, 076405 (2006).
\bibitem{oguri} D. Pines and P. Nozi\'eres. {\em ``The Theory of Quantum Liquids,''} Benjamin, New York, 1966; A. Oguri, Phys. Rev. B {\bf 64}, 153305 (2001).


\end{thebibliography}
\end{document}